\documentclass[aps,pra,reprint,showpacs,nofootinbib,superscriptaddress]{revtex4-1}
\usepackage{graphicx}
\usepackage{hyperref}
\usepackage{amsfonts}
\usepackage{amsmath,amssymb}
\usepackage{bm}
\usepackage{color}
\usepackage{epstopdf}
\usepackage{epsfig}
\usepackage{subfig}
\usepackage{float}
\usepackage{verbatim}
\usepackage{hyperref}
\hypersetup{
     colorlinks   = true,
     citecolor    = blue
}
\usepackage{lineno}
\usepackage{ulem}
\usepackage[thicklines]{cancel}

{\rm }

\def\be{\begin{equation}}
 \def\ee{\end{equation}}
 \def\bea{\begin{eqnarray}}
 \def\eea{\end{eqnarray}}

\def\2{\frac{1}{2}}
\def\4{\frac{1}{4}}

\catcode`\@=11

\def\@normalsize{\@setsize\normalsize{15pt}\xiipt\@xiipt
\abovedisplayskip 14pt plus3pt minus3pt%
\belowdisplayskip \abovedisplayskip
\abovedisplayshortskip  \z@ plus3pt%
\belowdisplayshortskip  7pt plus3.5pt minus0pt}
\def\small{\@setsize\small{13.6pt}\xipt\@xipt
\abovedisplayskip 13pt plus3pt minus3pt%
\belowdisplayskip \abovedisplayskip
\abovedisplayshortskip  \z@ plus3pt%
\belowdisplayshortskip  7pt plus3.5pt minus0pt
\def\@listi{\parsep 4.5pt plus 2pt minus 1pt
            \itemsep \parsep
            \topsep 9pt plus 3pt minus 3pt}}

\def\underline#1{\relax\ifmmode\@@underline#1\else
        $\@@underline{\hbox{#1}}$\relax\fi}
\@twosidetrue \relax

\catcode`@=12

\catcode`\@=11

\def\section{\@startsection{section}{1}{\z@}{3.5ex plus 1ex minus
   .2ex}{2.3ex plus .2ex}{\large\bf}}

\def\ps@headings{\def\@oddfoot{}\def\@evenfoot{}
\def\@oddhead{\hbox{}\hfill
        \makebox[.5\textwidth]{\raggedright\ignorespaces --\thepage{}--
        \hfill }}
\def\@evenhead{\@oddhead}
\def\subsectionmark##1{\markboth{##1}{}}
}

\ps@headings

\catcode`\@=12

\begin{document}

\title{Quantum Computation of Scattering Amplitudes in Scalar Quantum Electrodynamics
}

\author{K\"ubra Yeter-Aydeniz }
\email{kyeteraydeniz@tntech.edu}
\affiliation{Department of Physics, Tennessee Technological University,  Cookeville, TN 38505, USA}

\author{George Siopsis }
\email{siopsis@tennessee.edu}
\affiliation{Department of Physics and Astronomy,
	The University of Tennessee, Knoxville, TN 37996-1200, USA}

\date{\today}

\begin{abstract}

We present a quantum algorithm for the calculation of scattering amplitudes of massive charged scalar particles in scalar quantum electrodynamics. Our algorithm is based on continuous-variable quantum computing architecture resulting in exponential speedup over classical methods. We derive a simple form of the Hamiltonian including interactions, and a straightforward implementation of the constraint due to gauge invariance.

\end{abstract}

\maketitle

\onecolumngrid
\section{Introduction}
Quantum Field Theory (QFT) provides a unique perspective of the natural world, its composition and how it works, by combining the laws of quantum mechanics and special relativity. QFT also forms the foundation of the Standard Model of elementary particles which encompasses all known particles and interactions between them with the exception of gravity. Theoretical predictions are tested experimentally using scattering. The calculation of scattering amplitudes is a daunting task with classical computers. With the help of a perturbative expansion utilizing Feynman diagrams, calculations are possible at weak coupling. For strong coupling, lattice field theory is used which becomes exponentially harder as the number of sites grows \cite{Jordan:2011ci}.

There are two ways to process information in a quantum system. One way is to encode the information into discrete systems, like the spin of an electron or the polarization of a photon. Quantum algorithms are implemented on quantum systems with discrete spectra. Another way to encode the information is to use continuous variables (CV), which was first proposed by Lloyd and Braunstein \cite{Lloyd1999}, relying on quantum systems whose observables have continuous spectra, like the position and momentum of a particle or the quadratures of an electromagnetic field \cite{Pati2000cv,Lomonaco2003,Braunstein2003,Adcock2009, Zwierz2010,Braunstein1999,  Li2002, Braunstein2003,Laudenbachs2017, Grosshans2003,Aoki2009,Vaidman1994, Braunstein1998} (also see \cite{Andersen2015} for a discussion of hybrid schemes leading to quantum algorithms that benefit from the advantages of both discrete and continuous variables). 
%
%
%
The experimental advantages brought on by CV quantum architectures have resulted in an extensive use of CVs in recent years. For instance, the extension of the discrete formalism of cluster state protocols to the continuous formalism \cite{Menucci2006} provided the advantage of the deterministic production of multipartite entangled states, and measurement of high fidelity, using present technology \cite{Lau2013}. The physical realization of CV cluster-state quantum computing has led to proof-of-principle experimental demonstrations such as a fully tunable gate for continuous-variable one-way quantum computation \cite{Yokoyama2015}, a dynamical squeezing gate for universal quantum information processing \cite{Miyata2014}, and large entangled states for scalable quantum information and quantum computing \cite{Pysher2011}.


The purpose of this paper is to present a quantum calculation of scattering amplitudes of a massive charged scalar particle in quantum electrodynamics (QED). To this end, we will utilize a CV quantum architecture as a tool to describe our quantum system, since it has certain advantages compared to discrete variables. Quantum algorithms for the calculation of scattering amplitudes give an exponential speedup compared to any known classical algorithms. This was first shown by Jordan, \textit{et al.}, who introduced a quantum algorithm using discrete variables for the simulation of a scalar bosonic QFT \cite{Jordan:2011ci}. Marshall, \textit{et al.}, adapted the results of ref.\ \cite{Jordan:2011ci} to the case of a CV quantum architecture \cite{Marshall:2015mna}. Others have studied the quantum simulation of QED in order to understand the elementary constituents of matter, using engineered and controlled quantum mechanical devices, such as trapped ions \cite{Martinez2016, Yang:2016hjn}, or lattice gauge models \cite{Notarnicola:2015sia, Ercolessi:2017jbi}.
We will follow the method developed in \cite{Marshall:2015mna}, and introduce a CV quantum algorithm for the calculation of scattering amplitudes of massive charged scalars in QED. 

In our approach, in addition to extending the CV algorithms for the computation of scattering amplitudes of a scalar QED system, we introduce a straightforward implementation of the constraint which is present due to the gauge invariance of the system. Gauge invariance leads to photons having two degrees of freedom (transverse polarization), even though they are described by a four-vector (gauge field). A reduction to the two physical degrees of freedom is usually accomplished by imposing a gauge-fixing condition. We will follow the alternative approach of augmenting the Lagrangian with the addition of a term that leads to a non-singular propagator. This leads to a photonic system that uses more than two fields for its description, the additional fields corresponding to unphysical degrees of freedom. Consequently, the Hilbert space contains unphysical states. They are seen to decouple by imposing Gauss's Law, which is a constraint on the system, and generates gauge invariance. In the calculation of scattering amplitudes, we turn the coupling (electric charge) on and off adiabatically. This breaks gauge invariance involving time-dependent transformations, and one ought to exercise care in imposing Gauss's Law. We show how this can be done in a straightforward manner in the quantum algorithm.

Our discussion is organized as follows. In section \ref{sec:set}, we discuss our scalar quantum electrodynamic system. We introduce the fields and the Hamiltonian defined on a lattice. We discuss the complications due to gauge invariance and renormalization, and introduce a straightforward method to account for the constraint (Gauss's Law). In section \ref{sec:QC}, we discuss quantum computation of the initial state, scattering amplitudes, and measurement of the final state. Finally, in section \ref{sec:con}, we provide a summary and outlook.

\section{The system}
\label{sec:set}

\subsection{Non-interacting scalars and photons}
Ignoring interactions at the moment, our system consists of two massive scalar fields $\phi_1,\phi_2$, both of mass $m$, and a four-vector (gauge) field $(A_t, \bm{A})$ representing photons.
The Hamiltonian for each real scalar field in three spatial dimensions is \cite{Marshall:2015mna}
\be H_{\phi_i} = \frac{1}{2} \int_{[0,L]^3} d^3x  \left[ \pi_i^2 -\phi_i {\bm{\nabla}}^2 \phi_i + m^2 \phi_i^2 \right] \ee
for $i=1,2$, where $L$ is the size of each spatial dimension, and we impose periodic boundary conditions. $\pi_i$ is the conjugate momentum, obeying commutation relations
\be\label{eq1} [\phi_i (\bm{x}) , \pi_j (\bm{x}')] = i\delta_{ij} \delta^3 (\bm{x} - \bm{x}') ~.\ee
In the presence of interactions, the mass parameter, as well as the fields are renormalized due to quantum effects. This effect is often described by the term \textit{bare} referring to the parameters and fields of the classical Lagrangian, and \textit{dressed} referring to corrected parameters and fields due to quantum effects. Here, we will not be concerned with the renormalization of fields which are treated as dynamical variables (in a path-integral formulation, one integrates over them), but quantum corrections to (renormalization of) the mass parameter are important.
	In our approach, we will keep $m$ as close to the physical (\textit{dressed}, renormalized) mass as possible by introducing a counter term (correction mass term in the Hamiltonian). In the weak-coupling limit, this is achieved by an analytic calculation of renormalization using perturbation theory, as we explain later on. In the strong-coupling regime, the physical mass parameter can only be determined in retrospect, by calculating the poles of correlators (Green functions).

We may expand the fields in creation and annihilation operators as
\be \phi_i({\bm{x}})= \frac{1}{L^{3/2}}\sum_{\bm{k}}\frac{1}{\sqrt{2\omega(\bm{k})}}\left( a_i (\bm{k}) e^{i{\bm{k}} \cdot {\bm{x}}}+a_i^\dagger (\bm{k}) e^{-i{\bm{k}} \cdot {\bm{x}}} \right) \ , \ \ \pi_i({\bm{x}})= \frac{i}{L^{3/2}}\sum_{\bm{k}}\sqrt{\frac{\omega(\bm{k})}{2}} \left( -a_i (\bm{k}) e^{i{\bm{k}} \cdot {\bm{x}}}+a_i^\dagger (\bm{k}) e^{-i{\bm{k}} \cdot {\bm{x}}} \right)~.
\ee
where $\frac{L}{2\pi} \bm{k} \in \mathbb{Z}^3$, and $\omega(\bm{k})=\sqrt{{\bm{k}}^2+m^2}$, and we used the time evolution of the scalar fields, $\partial_t \phi_i = \pi_i$.
The commutation relations \eqref{eq1} are easily seen to reduce to standard commutation relations of the Fourier modes,
\be
[a_i(\bm{k}), a_j^\dagger(\bm{k'})] =\delta_{ij}\delta_{{\bm{k}}{\bm{k'}}}~.
\ee
It is convenient to combine the two real scalar fields into
a complex scalar field $\phi = \frac{1}{\sqrt{2}} \left(\phi_1+i\phi_2\right)$ and its conjugate momentum $\pi = \frac{1}{\sqrt{2}} \left(\pi_1-i\pi_2\right)$. The complex field can be expanded into creation and annihilation operators as
\be
\phi({\bm{x}})= \frac{1}{L^{3/2}}\sum_{\bm{k}}\frac{1}{\sqrt{2\omega(\bm{k})}} \left( b (\bm{k}) e^{i{\bm{k}} \cdot {\bm{x}}}+c^\dagger (\bm{k}) e^{-i{\bm{k}} \cdot {\bm{x}}} \right)~,
\ee
with $b = \frac{a_1+ia_2}{\sqrt{2}}, c = \frac{a_1-ia_2}{\sqrt{2}}$.  The two creation operators, $b^\dagger (\bm{k})$ and $c^\dagger (\bm{k})$, create particles and anti-particles, respectively. The normal-ordered Hamiltonian for the complex scalar field reads
\begin{equation}
H_\phi= H_{\phi_1} + H_{\phi_2} = \sum_{{\bm{k}}} \omega(\bm{k}) \left( b^\dagger (\bm{k}) b(\bm{k})+c^\dagger (\bm{k}) c (\bm{k}) \right)~.
\end{equation}
Another useful quantity is the scalar Green function $G_\phi (\bm{x}, t)$ which satisfies
\be \left( \partial_t^2 - {\bm{\nabla}}^2 + m^2 \right) G_\phi (t, \bm{x}) = \delta (t)\delta^3 (\bm{x}) ~. \ee
By taking Fourier transform, we easily obtain in the large-$L$ limit,
\be\label{eqGphi} G_\phi (t, \bm{x}) = \int \frac{dE}{2\pi} \int \frac{d^3p}{(2\pi)^3} e^{i(Et- \bm{p}\cdot \bm{x})} \tilde{G}_\phi (E,\bm{p}) \ , \ \ \tilde{G}_\phi (E,\bm{p}) = \frac{i}{E^2 - \bm{p}^2 - m^2}~.\ee
Turning to photons, quantization is not straightforward due to gauge invariance (constrained system). The Lagrangian is $L = \frac{1}{2} \int d^3 x (\bm{E}^2 - \bm{B}^2)$, where $\bm{E} = -\bm{\nabla} A_t - \partial_t \bm{A}$ is the electric field and $\bm{B} = \bm{\nabla} \times \bm{A}$ is the magnetic field. Even though the system is described by a four-component gauge field $(A_t, \bm{A})$, it contains only two physical degrees of freedom (transverse polarization). To reveal them, one usually imposes a gauge-fixing condition. Instead, for our purposes, we find it more
convenient to add the term $-\frac{\lambda}{2} \int d^3 x (\partial_t A_t + \bm{\nabla} \cdot \bm{A})^2$, where $\lambda$ is an arbitrary parameter that should not affect the physics. We obtain the conjugate momenta $\varpi_t = -\lambda (\partial_t A_t + \bm{\nabla} \cdot \bm{A})$, and $\bm{\varpi} = - \bm{E}$. Notice that $\varpi_t = 0$, if $\lambda =0$, showing that our system is constrained.
After standard steps, we arrive at the Hamiltonian
\begin{equation}
H_{\text{QED}}=\frac{1}{2}\int_{[0,L]^3} d^3x  \left[-\frac{1}{\lambda}(\varpi_t+\lambda \bm{\nabla}\cdot \bm{A})^2+ \bm{\varpi}^2 -\bm{A}\cdot {\bm{\nabla}}^2 \bm{A} + (\lambda-1) (\bm{\nabla}\cdot \bm{A})^2+ 2A_t \bm{\nabla}\cdot  \bm{\varpi} \right]~. \label{QED_H}
\end{equation}
Notice that $A_t$ acts as a Lagrange multiplier imposing the constraint (Gauss's Law)
\be\label{eq10} \bm{\nabla}\cdot  \bm{\varpi} = 0 \ee
and its conjugate momentum decouples ($A_t=0$ gauge). Moreover, since no physical results depend on $\lambda$, we will make the convenient choice $\lambda =1$ (Feynman gauge). The Hamiltonian simplifies to
\begin{equation}
H_{\text{QED}}=\frac{1}{2}\int_{[0,L]^3} d^3x \left[ \bm{\varpi}^2 -\bm{A}\cdot {{\nabla}}^2 \bm{A}  \right] \label{QED_H1}
\end{equation}
described in terms of a three-component field which includes an unphysical degree of freedom. It will be eliminated after imposing the constraint \eqref{eq10}. Once interactions are turned on, the gauge field receives quantum corrections (is renormalized), so the field appearing in the classical Lagrangian is \textit{bare}. However, we will not need to be concerned about field renormalization here, because the fields are treated as dynamical variables (integrated over in a path-integral formulation).

To quantize the system, we impose standard commutation relations on the gauge potential $\bm{A}$ and its conjugate momentum (electric field) $\bm{\varpi}$,
\begin{equation}
[A_i({\bm{x}}), \varpi_j({\bm{x}}')]=i\delta_{i j} \delta^3({{\bm{x}}-{\bm{x}}'})~.
\end{equation}
The gauge potential can be expanded in a manner similar to the scalar field,
\begin{equation}\label{eq13}
\bm{A}({\bm{x}})= \frac{1}{L^{3/2}} \sum_{{\bm{k}}} \frac{1}{\sqrt{2\omega_\gamma(\bm{k})}}\left[\bm{a}({\bm{k}})e^{i {\bm{k}}\cdot {\bm{x}}}+\bm{a}^\dagger({\bm{k}})e^{-i {\bm{k}}\cdot {\bm{x}}}\right]~,
\end{equation}
where $\omega_\gamma (\bm{k}) = |\bm{k}|$. We deduce
the commutation relations
\be [a_i (\bm{k}) , a_j^\dagger (\bm{k}')] = \delta_{ij} \delta_{\bm{k}  \bm{k}'}~.
\ee
The normal-ordered Hamiltonian reads
\be H_{\text{QED}} = \sum_{{\bm{k}}} \omega_\gamma (\bm{k}) \bm{a}^\dagger (\bm{k}) \cdot \bm{a} (\bm{k}) ~. \ee
Unphysical states are rejected by demanding that physical states  obey the quantum extension of Gauss's Law \eqref{eq10}. To implement it, it is convenient to consider the commutator of \eqref{eq10} with the Hamiltonian \eqref{QED_H1}. We easily obtain $\bm{\nabla}\cdot \bm{A} =0$. After Fourier transforming to momentum space using \eqref{eq13}, at the quantum level, we impose the gauge condition
\be\label{eqGa} \bm{k}\cdot \bm{a} (\bm{k}) |\Psi\rangle = 0 ~. \ee
Evidently, the ground state obeys the constraint \eqref{eqGa}. It is easily seen that a single-photon state $\bm{\zeta}\cdot \bm{a}^\dagger (\bm{k})  |0\rangle$ obeys \eqref{eqGa} provided its polarization is transverse, $\bm
k \cdot \bm{\zeta} = 0$. Transversality is a general condition for states containing an arbitrary number of photons to obey the gauge condition \eqref{eqGa}. This shows that unphysical states do indeed decouple and the system has only two degrees of freedom (transverse polarization), even though it is described in terms of a three-component gauge field $\bm{A}$.

As with the scalar (eq.\ \eqref{eqGphi}), we introduce the photon propagator
\be\label{eqGgamma} G_\gamma (t, \bm{x}) = \int \frac{dE}{2\pi} \int \frac{d^3p}{(2\pi)^3} e^{i(Et- \bm{p}\cdot \bm{x})} \tilde{G}_\gamma (E,\bm{p})\ , \ \ \tilde{G}_\gamma (E,\bm{p}) = \frac{i}{E^2 - \bm{p}^2}~, \ee
showing that the photon is massless. The vanishing mass should receive no quantum corrections. This is ensured by gauge invariance at the quantum level.

For numerical calculations, we shall discretize space, thus putting the system on a lattice. Let $a$ be the lattice spacing. Then in each spatial dimension we have $L/a$ points. We will set $a=1$ so that $L$ is an integer ($x_i = 0,1,\dots, L-1$, and $L\gg 1$). Notice that in these units, the continuum limit, which would normally be taken as $a\to 0$, is instead the limit in which all physical quantities of positive length dimension become very large, e.g., $L\to\infty$. The momentum $\bm{k}$ lives on the dual lattice ($\frac{L}{2\pi} k_i = 0,1,\dots, L-1$). Our system consists of $5L^3$ harmonic oscillators ($2L^3$ describing the complex scalar field and $3L^3$ describing the gauge field \footnote{{Recall that, even though the photon has two degrees of freedom, our system is described by a three-component gauge field $\bm{A}$, and the unphysical degree of freedom is dealt with by making sure that unphysical states decouple.}}). 
The scalar modes can be written in terms of the scalar field as
\be\label{eq20} b(\bm{k}) =   \sqrt{\frac{\omega(\bm{k})}{2}} \tilde{\phi} (\bm{k}) + \frac{i}{\sqrt{2\omega (\bm{k})}} \tilde{\pi}^\dagger (\bm{k} ) \ , \ \ c(\bm{k}) =  \sqrt{\frac{\omega(\bm{k})}{2}} \tilde{\phi}^\dagger (\bm{k}) + \frac{i}{\sqrt{2\omega (\bm{k})}} \tilde{\pi} (\bm{k} )  \ee
and the photon modes in terms of the gauge field are
\be\label{eq21} a_i(\bm{k}) =  \sqrt{\frac{\omega_\gamma (\bm{k})}{2}} \tilde{A}_i (\bm{k}) + \frac{i}{\sqrt{2\omega_\gamma (\bm{k})}} \tilde{\varpi}_i (\bm{k} )  \ , \ \ i=1, 2 , 3~, \ee
where we introduced the Fourier transform of the field $\phi (\bm{x})$,
\be\label{eqFT} \tilde{\phi} (\bm{k}) \equiv \frac{1}{L^{3/2}} \sum_{\bm{x}} e^{-i\bm{k}\cdot \bm{x}}  \phi (\bm{x}) \ee
and similarly for the other fields, and we defined
\be \omega^2 (\bm{k}) = m^2 + 4 \sum_{i=1}^3 \sin^2 \frac{k_i}{2} \ , \ \ \omega_\gamma (\bm{k}) = \omega (\bm{k}) |_{m = \frac{1}{L}}~. \ee
Notice that we set the mass of the photon to $\frac{1}{L} \ll 1$, and not zero, in order to avoid numerical problems with zero modes.

The non-interacting Hamiltonian reads
\be\label{eqH0} H_0 = H_\phi + H_{\text{QED}} = \sum_{{\bm{k}}} \omega({\bm{k}}) \left( b^\dagger (\bm{k}) b(\bm{k}) + c^\dagger (\bm{k}) c(\bm{k}) \right) + \sum_{{\bm{k}}} \omega_\gamma (\bm{k}) \bm{a}^\dagger (\bm{k}) \cdot \bm{a} (\bm{k}) ~. \ee

The scalar and photon propagators, eqs.\ \eqref{eqGphi} and \eqref{eqGgamma}, respectively, in the large-$L$ limit turn into
\bea\label{eqGphi2} G_\phi (t, \bm{x}) &=& \int \frac{dE}{2\pi} \int_{[0,2\pi]^3} \frac{d^3p}{(2\pi)^3} e^{i(Et- \bm{p}\cdot \bm{x})} \tilde{G}_\phi (E,\bm{p})\ , \ \ \tilde{G}_\phi (E,\bm{p}) = \frac{i}{E^2 - \omega^2 (\bm{p}) }\ , \nonumber\\ G_\gamma (t, \bm{x}) &=& \int \frac{dE}{2\pi} \int_{[0,2\pi]^3} \frac{d^3p}{(2\pi)^3} e^{i(Et- \bm{p}\cdot \bm{x})} \tilde{G}_\gamma (E,\bm{p})\ , \ \ \tilde{G}_\gamma (E,\bm{p}) = \frac{i}{E^2 - \omega_\gamma^2 (\bm{p}) } ~.\eea
These expressions will be useful in the calculation of Feynman diagrams when we include interactions.

\subsection{Interactions and Renormalization}
\label{sec:ren}

In our discussion so far, we considered only free particles (scalars and photons). This is a valid description of scalars with no electric charge. We now switch on interactions between the scalars and the photons by assigning an electric charge $e$ to the scalar particles. The Hamiltonian is modified by the addition of the interaction Hamiltonian 
\begin{equation}
H_{I}(e)=-\sum_{{\bm{x}}} \left[ ie\bm{A} \cdot (\phi \bm{\nabla}\phi^\dagger-\phi^\dagger \bm{\nabla}\phi)+e^2 \bm{A}^2\phi^\dagger \phi \right] \label{Hint}
\end{equation}
where the gradients represent finite differences over neighboring points on the lattice. 
Quantization of the system yields quantum corrections (renormalization) of the fields and the parameters ($m$ and $e$). As explained above, we will not be concerned with field renormalization. Mass renormalization necessitates the addition of a mass counter term to the Hamiltonian,
\begin{equation}\label{eqHct}
H_{\text{c.t.}}=\frac{\delta m}{2}\sum_{{\bm{x}}}  \phi^\dagger \phi
\end{equation}
with the parameter $\delta m$ to be determined. As we switch the coupling $e$ on and off, we also need to switch $\delta m$ in tandem. The counter term modifies the effective mass to \be\label{eqm0} m_0^2 = m^2 + \delta m~. \ee
Ideally, we will choose $\delta m$ so that the effective mass $m_0$ is the bare mass and $m$ is the physical (renormalized, dressed) mass of the charged scalar field. This is not always possible. In general, the physical mass will not coincide with $m$; it is found by calculating the poles of a Green function. In the weak-coupling limit, which is usually the physically relevant regime, the bare mass $m_0$ can be estimated by perturbation theory (given an experimentally observed value of $m$) to a high degree of accuracy.

Thus, the full Hamiltonian of our system is
\be\label{eqH} H = H_{0} + H_I(e)+ H_{\text{c.t.}} ~. \ee
Gauss's Law \eqref{eq10} is also modified to
\be\label{eqGi} \bm{\nabla}\cdot \bm{\varpi} - \rho  = 0 \ \ , \ \ \rho = ie(\pi \phi -\pi^\ast \phi^\ast  ) \ee
It generates the gauge transformation
\be \bm{A} \to \bm{A} + \bm{\nabla} \chi \ , \ \ \phi \to e^{ie\chi} \phi \ , \ \ \pi \to e^{-ie\chi} \pi\ee
where $\chi (\bm{x})$ is an arbitrary time-independent function. The part of the Hamiltonian that involves the scalar field ($H-H_{\text{QED}}$) is invariant under this transformation. Notice that this is true even if the coupling constant is time dependent ($e=e(t)$), which is the case as it is being adiabatically switched on or off in the calculation of scattering amplitudes. It follows that the part of the Hamiltonian that has a non-vanishing commutator with Gauss's Law \eqref{eqGi} is the electromagnetic part, $H_{\text{QED}}$, and so the commutator of Gauss's Law with the Hamiltonian leads to the same constraint as in the non-interacting case considered above, $\bm{\nabla}\cdot \bm{A} = 0$. Consequently, the quantum extension of Gauss's Law, i.e., the gauge condition \eqref{eqGa} we imposed above, remains the same after interactions are switched on. It suffices to prepare a physical initial state (i.e., one  obeying the constraint \eqref{eqGa}) for the state to remain physical through evolution under the full interacting Hamiltonian \eqref{eqH}.

\begin{figure}
	\begin{center}
		\includegraphics[width=0.5\textwidth]{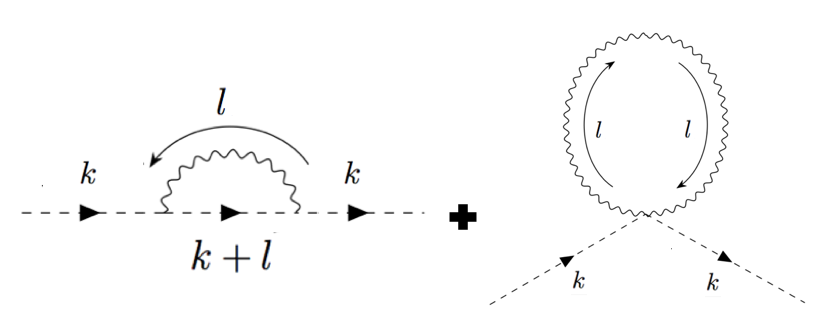}
	\end{center}
	\caption {\label{scalaroneloop} One-loop contributions to the scalar mass renormalization. }
\end{figure}

Having determined the implementation of the gauge constraint, we turn to the determination of the mass counter term parameter $\delta m$. As noted above, it is desirable to set its value so that $m_0$ (eq.\ \eqref{eqm0}) is the bare mass of the complex scalar so that $m$ is the physical mass. It can be calculated analytically in the case of weak coupling by using perturbation theory.
At lowest perturbative order, the Feynman diagrams contributing to $\delta m$ are shown in figure \ref{scalaroneloop}.
We obtain the contributions to the scalar self-energy, respectively,
\bea \Sigma^{(1)}_\phi (k^0,\bm{k}) &=& -2e^2 \int_{-\infty}^{\infty} \frac{dl^0}{2\pi} \int_{[-\pi,\pi]^3} \frac{d^3l}{(2\pi)^3} \left( (l^0+2k^0)^2 - (\bm{l}+2\bm{k})^2 \right) \tilde{G}_\gamma (l^0, \bm{l}) \tilde{G}_\phi (l^0+k^0,\bm{l}+ \bm{k}) ~,\nonumber\\
\Sigma^{(2)}_\phi (k^0,\bm{k}) &=& 8ie^2 \int_{-\infty}^{\infty}  \frac{dl^0}{2\pi} \int_{[-\pi,\pi]^3} \frac{d^3l}{(2\pi)^3} \tilde{G}_\gamma (l^0, \bm{l}) ~,\eea
where $\tilde{G}_\phi$, and $\tilde{G}_\gamma$ are defined in eq.\ \eqref{eqGphi2}.
The mass counter term parameter is given by
\be \delta m = \left. \left( \Sigma^{(1)}_\phi (k^0,\bm{k}) + \Sigma^{(2)}_\phi (k^0,\bm{k}) \right) \right|_{(k^0)^2 = \bm{k}^2 + m^2} + \mathcal{O} (e^4)~. \ee
We are interested in the limit in which the lattice spacing $a\to 0$. Since we are using units in which $a=1$, in this limit, quantities with the dimension of energy do not contribute at leading order. After setting $k^0, \bm{k}, m$ to zero, we obtain the expression for the mass counter term
\be \delta m = -6e^2 \int_{-\infty}^{\infty}  \frac{dl^0}{2\pi} \int_{[-\pi,\pi]^3} \frac{d^3l}{(2\pi)^3} \frac{1}{(l^0)^2 - 4 \sum_{i=1}^3 \sin^2 \frac{l^i}{2}} + \mathcal{O} (e^4) ~.\ee
The integrals can be evaluated numerically. We obtain
\be\label{eqme} \delta m = \left[ - 1.36 + \mathcal{O} (m^2) \right] e^2 + \mathcal{O} (e^4) \ee
which is valid in the limit $a\to 0$.\footnote{Recall that we have set $a=1$. Were we to restore the lattice spacing $a$, the leading-order contribution to $\delta m$ would be $-\frac{1.36}{a^2} e^2$, which diverges in the continuum limit $a\to 0$.} A better approximation to $\delta m$ can be achieved by including higher-order perturbative corrections, as long as the coupling constant is small. Physically, $\frac{e^2}{4\pi}$ is the fine-structure constant $\frac{1}{137}$, which is a small parameter. If $e$ is not small (strong-coupling regime), then $m$ and $\delta m$ can be adjusted by making sure that the mass pole of correlation functions is the physical mass of the complex scalar.

\begin{figure}
	\begin{center}
		\includegraphics[width=0.5\textwidth]{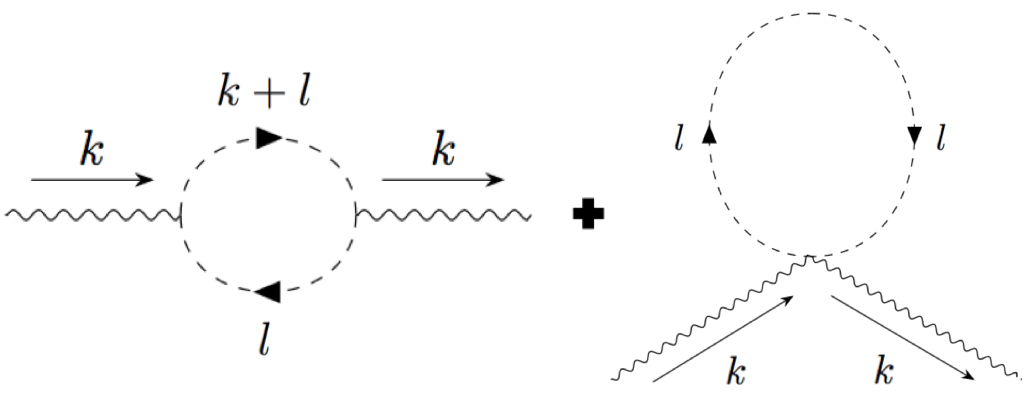}
	\end{center}
	\caption {\label{photononeloop} One-loop contributions to photon field (charge) renormalization. }
\end{figure}

The electric charge $e$ is also renormalized by the interactions. Because of gauge invariance, this renormalization can be deduced from the photon propagator. At the lowest order, the diagrams that contribute are shown in figure \ref{photononeloop}. We obtain the contributions to the vacuum photon polarization, respectively,
\bea
\Pi^{(1)} (k^0, \bm{k}; \bm{\zeta}_1, \bm{\zeta}_2)&=& -4e^2 \int_{-\infty}^{\infty}  \frac{dl^0}{2\pi} \int_{[-\pi,\pi]^3} \frac{d^3l}{(2\pi)^3} \bm{\zeta}_1\cdot \bm{l}\bm{\zeta}_2\cdot \bm{l} \tilde{G}_\phi (l^0, \bm{l}) \tilde{G}_\phi (l^0+k^0,\bm{l}+ \bm{k})~,\nonumber\\
\Pi^{(2)}  (k^0, \bm{k}; \bm{\zeta}_1, \bm{\zeta}_2) &=& 
2e^2 \bm{\zeta}_1 \cdot \bm{\zeta}_2\int_{-\infty}^{\infty}  \frac{dl^0}{2\pi} \int_{[-\pi,\pi]^3} \frac{d^3l}{(2\pi)^3} \tilde{G}_\phi (l^0, \bm{l}) \eea
where $\bm{\zeta}_{1,2}$ are the photon polarizations of the two external legs ($\bm{\zeta}_{1,2} \cdot \bm{k} = 0$). 
The calculation of $\Pi^{(1)}$ proceeds as follows. We introduce a Feynman parameter to write the propagators as
\be \tilde{G}_\phi (l^0, \bm{l}) \tilde{G}_\phi (l^0+k^0,\bm{l}+ \bm{k}) = -\int_0^1 dx \frac{1}{[(l^0 + (1-x)k^0)^2 + x(1-x) (k^0)^2 - x\omega (\bm{l}) - (1-x) \omega (\bm{l} + \bm{k}) ]^2}~.\ee
After shifting $l^0 \to l^0 - (1-x)k^0$, the integral over $l^0$ is easily performed. We obtain
\be \Pi^{(1)} (k^0, \bm{k}; \bm{\zeta}_1, \bm{\zeta}_2) = 2e^2 \int_0^1 dx  \int_{[-\pi,\pi]^3} \frac{d^3l}{(2\pi)^3} \frac{\bm{\zeta}_1\cdot \bm{l}\bm{\zeta}_2\cdot \bm{l}}{[x\omega (\bm{l}) + (1-x) \omega (\bm{l} + \bm{k})-x(1-x) (k^0)^2 ]^{1/2}}~. \ee
Because of rotational invariance, and transversality, $\bm{\zeta}_{1,2}\cdot \bm{k} = 0$, the above expression simplifies to
\be \Pi^{(1)} (k^0, \bm{k}; \bm{\zeta}_1, \bm{\zeta}_2) = 2e^2 \bm{\zeta}_1\cdot\bm{\zeta}_2 \int_0^1 dx  \int_{[-\pi,\pi]^3} \frac{d^3l}{(2\pi)^3} \frac{ \bm{l}^2}{[x\omega (\bm{l}) + (1-x) \omega (\bm{l} + \bm{k})-x(1-x) (k^0)^2 ]^{1/2}}~. \ee
Expanding in $(k^0,\bm{k})$, we obtain
\be \Pi^{(1)} (k^0, \bm{k}; \bm{\zeta}_1, \bm{\zeta}_2) = e^2 \bm{\zeta}_1\cdot\bm{\zeta}_2 \left[ \Pi_0^{(1)}+ (k^0)^2 \Pi_1^{(1)}  + \bm{k}^2 \Pi_2^{(1)}  \dots \right] \ee
where $\Pi_i^{(1)}$ are constants that can be evaluated numerically in the limit $a\to 0$ (in which we can set $ma=0$). We obtain
\be \Pi_0^{(1)} = 0.455 +\mathcal{O} (m^2)\ , \ \ \Pi_1^{(1)} = -\Pi_2^{(1)} = \frac{1}{48\pi^2} \log \frac{1}{m^2} + 0.003 + \mathcal{O} (m^2)~. \ee
If we restore the lattice spacing $a$, by dimensional analysis the leading terms in the above expansions are $\frac{0.455}{a^2}$ and $\frac{1}{48\pi^2} \log \frac{1}{m^2a^2}$, which diverge in the continuum limit $a\to 0$.
	
Similarly, we obtain for the second Feynman diagram,
\be \Pi^{(2)}  (k^0, \bm{k}; \bm{\zeta}_1, \bm{\zeta}_2) = 
-e^2 \bm{\zeta}_1 \cdot \bm{\zeta}_2 \Pi_0^{(1)}\ee
and for the sum of the two diagrams,
\be \Pi^{(1)} (k^0, \bm{k}; \bm{\zeta}_1, \bm{\zeta}_2) + \Pi^{(2)}  (k^0, \bm{k}; \bm{\zeta}_1, \bm{\zeta}_2) = \left[\frac{1}{48\pi^2} \log \frac{1}{m^2} + 0.003\right] e^2 [(k^0)^2 - \bm{k}^2 ] \bm{\zeta}_1\cdot \bm{\zeta}_2 + \dots~.\ee
This implies photon field renormalization, and by gauge invariance, charge renormalization, with the bare electric charge given by
\be\label{eq45} e_0^2 = e^2 + \delta e \ , \ \ \delta e = \left[\frac{1}{48\pi^2} \log \frac{1}{m^2} + 0.003\right] e^4 + \dots~.\ee
Therefore, we ought to add a counter term to the Hamiltonian, as we did with the mass parameter. Alternatively, we ought to make sure that the coupling constant in the Hamiltonian is the bare electric charge $e_0$. Even though $\delta e$ diverges in the continuum limit, as remarked above, the divergence is logarithmic, and in practice the numerical value of the charge counter term $\delta e$ is small, so we will not include it explicitly here.

\section{Quantum computation} \label{sec:QC}

As explained in the previous section, we need $5L^3$ harmonic oscillators to build our system. $2L^3$ harmonic oscillators will be represented by the scalar conjugate variables $(\phi_i (\bm{x}), \pi_i(\bm{x}))$ ($i=1,2$), and $3L^3$ harmonic oscillators will be represented by the photon conjugate variables $(A_j (\bm{y}), \varpi_j (\bm{y}))$ ($j=1,2,3$). Due to its transverse polarization, the photon only has two degrees of freedom. In our formulation, the photon is described by a three-component gauge field. The redundant unphysical degree of freedom is eliminated in the Hilbert space by imposing the gauge condition \eqref{eqGa}. The harmonic oscillators will be represented by a continuous-variable quantum system consisting of $5L^3$ qumodes \cite{Weedbrook2012}.

The ground state $|0\rangle$ of the system is annihilated by all (a total of ${5}L^3$) annihilation operators, $\mathcal{B} (\bm{x}) = \frac{1}{\sqrt{2}} \left( \phi (\bm{x}) + i\pi^\dagger (\bm{x}) \right)$, $\mathcal{C} (\bm{x}) = \frac{1}{\sqrt{2}} \left( \phi^\dagger (\bm{x}) + i\pi (\bm{x}) \right)$,   and
$\mathcal{A}_i (\bm{x}) = \frac{1}{\sqrt{2}} \left( A_i (\bm{x}) + i\varpi_i (\bm{x}) \right)$ ($i=1,2{,3}$),
\be\label{eq43} \left\{ \mathcal{A}_i (\bm{x}) , \mathcal{B} (\bm{x}), \mathcal{C} (\bm{x}) \right\} |0\rangle = 0 ~. \ee
We will first build the initial state of the system without interactions, and then adiabatically switch on the coupling. Once the coupling reaches the desired value, we will let the system evolve under the Hamiltonian \eqref{eqH}. Subsequently, the coupling constant will be adiabatically switched off, and finally measurements will be performed on the free system.

\subsection{Initial State Preparation} \label{sec:initial}

Ground state preparation is one of the most challenging problems in quantum simulation \cite{Kitaev2002}; creating the ground state of our system is cumbersome as it involves a large number of modes.
The desired ground state $|\Omega\rangle$ of our system is annihilated by $b(\bm{k}), c(\bm{k}), a_i (\bm{k})$ ($i=1,2{,3}$), given by eqs.\ \eqref{eq20} and \eqref{eq21}, respectively,
\be\label{eq47} \left\{ a_i (\bm{k}) , b (\bm{k}), c (\bm{k}) \right\} |\Omega\rangle = 0~.  \ee
To build $|\Omega\rangle$, we will use eqs.\ \eqref{eq20} and \eqref{eq21} to construct a unitary transformation that relates $\{ a_i (\bm{k}) , b (\bm{k}), c (\bm{k}) \}$ to the modes $\{ \mathcal{A}_i (\bm{x}) , \mathcal{B} (\bm{x}), \mathcal{C} (\bm{x}) \}$. From eq.\ \eqref{eq20}, we obtain
\be b(\bm{k}) = \frac{1}{2}  \left( \sqrt{\omega (\bm{k})} + \frac{1}{\sqrt{\omega (\bm{k})}} \right) \tilde{\mathcal{B}} (\bm{k}) + \frac{1}{2} \left( \sqrt{\omega (\bm{k})} - \frac{1}{\sqrt{\omega (\bm{k})}} \right) \tilde{\mathcal{C}}^\dagger (\bm{k})  \ee
where $\tilde{\mathcal{B}} (\bm{k}) = \frac{1}{\sqrt{2}} \left( \tilde{\phi} (\bm{k}) + i\tilde{\pi}^\dagger (\bm{k}) \right)$ is the Fourier transform of $\mathcal{B} (\bm{x})$ (eq.\ \eqref{eqFT}), and similarly for $\tilde{\mathcal{C}} (\bm{k})$. These modes can be constructed in two steps. First, we apply the Fourier transform \eqref{eqFT} on $\mathcal{B} (\bm{x})$, which is a unitary transformation $\tilde{\mathcal{B}} = U_{\text{FT}} \mathcal{B}$, where the matrix elements of $ U_{\text{FT}}$ are given by
\be U_{\text{FT}} (\bm{k}, \bm{x}) = \frac{1}{L^{3/2}} e^{-i\bm{k}\cdot \bm{x}} ~.\ee
It follows that there is a Gaussian unitary $\mathcal{U}_{\text{FT}} (\mathcal{B})$ relating $\mathcal{B}$ to its Fourier transform $\tilde{\mathcal{B}}$,
\be \tilde{\mathcal{B}} = \mathcal{U}_{\text{FT}}^\dagger (\mathcal{B})\, \mathcal{B} \, \mathcal{U}_{\text{FT}} (\mathcal{B})~,\ee
involving only the modes $\mathcal{B} (\bm{x})$, that can be implemented with beam splitters and phase shifters  \cite{Reck1994}. Similarly, $\tilde{\mathcal{C}} (\bm{k})$ is constructed from the modes $\mathcal{C}$ via the unitary transformation $\tilde{\mathcal{C}} = \mathcal{U}_{\text{FT}}^\dagger (\mathcal{C})\, \mathcal{C} \, \mathcal{U}_{\text{FT}} (\mathcal{C})$. Having constructed the modes $\tilde{\mathcal{B}} (\bm{k})$ and $\tilde{\mathcal{C}} (\bm{k})$, the desired modes $b(\bm{k})$ are constructed by applying the two-mode down-converters \cite{Braunstein2005}
\be \mathcal{U}_2 (\bm{k}) = e^{\frac{\xi (\bm{k})}{2} \left( \tilde{\mathcal{B}}^\dagger (\bm{k}) \tilde{\mathcal{C}}^\dagger (\bm{k}) - \tilde{\mathcal{B}} (\bm{k})\tilde{\mathcal{C}} (\bm{k}) \right)} \ , \ \ e^{\xi (\bm{k})} = \omega (\bm{k})~.
\ee
We obtain
\be b(\bm{k}) = \mathcal{U}_2^\dagger (\bm{k}) \tilde{\mathcal{B}} (\bm{k})  \mathcal{U}_2 (\bm{k})~. \ee
The other modes are constructed similarly. Therefore, there exists a unitary $\mathcal{U}$, consisting of Gaussian operations that can be implemented with linear optical elements, that relates $\{ a_i (\bm{k}) , b (\bm{k}), c (\bm{k}) \}$ to the modes $\{ \mathcal{A}_i (\bm{x}) , \mathcal{B} (\bm{x}), \mathcal{C} (\bm{x}) \}$,
\be\label{eq53} b = \mathcal{U}^\dagger  \mathcal{B}   \mathcal{U} \ , \ \ c = \mathcal{U}^\dagger  \mathcal{C}   \mathcal{U} \ , \ \ a_i = \mathcal{U}^\dagger  \mathcal{A}_i   \mathcal{U} \ \ (i=1,2{,3})~.\ee
We can use the unitary $\mathcal{U}$ to construct the desired ground state as
\be |\Omega\rangle = \mathcal{U}^\dagger |0\rangle ~.\ee
Evidently, this state is annihilated by all modes $\{ a_i (\bm{k}) , b (\bm{k}), c (\bm{k}) \}$ on account of eqs.\ \eqref{eq43} and \eqref{eq53}, and is a physical state obeying the gauge condition \eqref{eqGa}.

For scattering amplitudes, we need to build excited states. We will consider scattering of scalar particles. Antiparticles can be treated similarly, but we will not consider photons in the initial state, because they are massless excitations and the adiabatic evolution would take a very long time. Adiabatic evolution of scalars takes $\mathcal{O} (1/m)$ time if they are self-interacting \cite{Jordan:2011ci}. If they are also interacting with an electromagnetic field, the time estimate of adiabatic evolution receives logarithmic corrections \cite{bibadia}.

A state containing a single scalar particle of momentum $\bm{k}$ is $b^\dagger (\bm{k}) |\Omega\rangle$. To construct it, we first create the single-mode excited state $\mathcal{B}^\dagger (\bm{x}) |0\rangle$, where $\bm{k} = \frac{2\pi}{L} \bm{x}$. There are various methods to construct a single-mode excited state \cite{Hong1986}. Despite considerable effort towards developing deterministic single-photon sources, pair-based heralded single-photon sources produced by parametric down conversion are still the most extensively used sources that can simulate excited states \cite{Migdall2013} (see also the method outlined in Appendix C of \cite{Marshall:2015mna}). Having constructed a single-mode excited state, we then apply the unitary $\mathcal{U}^\dagger$ to produce the single-particle state in our system, using
\be b^\dagger (\bm{k}) |\Omega\rangle = \mathcal{U}^\dagger \mathcal{B}^\dagger (\bm{x}) |0\rangle \ , \ \ \bm{k} = \frac{2\pi}{L} \bm{x}~. \ee
It is straightforward to extend this to an arbitrary number of excitations. Standard techniques also allow us to engineer wave packets from superpositions of single-mode excitations, $\sum_{{\bm{k}}} f(\bm{k}) b^\dagger (\bm{k}) |\Omega\rangle$, where the profile $f(\bm{k})$ is strongly peaked at a given momentum $\bm{k} = \bm{k}_0$, as needed for scattering amplitudes. Since all these excited states do not involve photon excitations, they trivially obey the gauge constraint \eqref{eqGa}, and hence they are physical.

\subsection{Scattering Amplitudes}
\label{sec:scat}
A scattering amplitude can be written as
\begin{equation}\label{eq56}
\mathcal{S}=\langle \text{out} | \mathcal{T} \, \exp\left\{ \int_{-T}^{T}dt \left[ H_{\text{int}}(t)+H_{\text{c.t.}}(t)\right]\right\}| \text{in} \rangle
\end{equation}
where $\mathcal{T}$ denotes time-ordering of operators, and we are interested in the limit $T \to \infty$. The incoming and outgoing states, $|\text{in}\rangle$ and $|\text{out}\rangle$, respectively, are states in the Hilbert space of the non-interacting Hamiltonian, $H_0$ (eq.\ \eqref{eqH0}). 

The initial state is constructed at time $t=-T$, as outlined above, to be the state of $N$ scalar particles,
\be |\text{in}\rangle = \prod_{n=1}^{N} \left( \sum_{{\bm{k}}} f_n (\bm{k})b^\dagger (\bm{k}) \right) |\Omega\rangle \ee
where the profile $f_n (\bm{k})$  is strongly peaked at momentum $\bm{k} = \bm{k}_n$.
It evolves in time with the application of a sequence of evolution operators in the interaction picture,
\begin{equation}
U(t)=e^{i t H_0}  e^{ i \delta t \left[H_{I} (e)+H_{\text{c.t.}}\right]}e^{-i t H_0}~,
\end{equation}
where $H_{I} (e)$ and $H_{\text{c.t.}}$ represent interactions and are given by eqs.\ \eqref{Hint} and \eqref{eqHct}, respectively. The initial state $|\text{in}\rangle$ trivially obeys the gauge constraint \eqref{eqGa} and is therefore a physical state. As discussed in the previous section, the constraint is not altered by the time evolution of the system, because of gauge invariance. Therefore, the evolved state $U(t)|\text{in}\rangle$ remains physical at all times $t$.
	
Using the Lie-Trotter product formula, we can approximate
the scattering amplitude \eqref{eq56} by
\begin{equation}
\mathcal{S}\approx \langle \text{out} |\left[ e^{i \delta t H_0} e^{i \delta t H_{I} (e)} e^{i\delta t H_{\text{c.t.}}}
\right]^{\frac{2 T}{\delta t}} | \text{in} \rangle
\end{equation}
where we  divided the time interval into $\frac{2 T}{\delta t}$ segments. The approximation becomes exact in the limit $\delta t\to 0$.

We divide the time interval $[-T,T]$ into three segments, $[-T,-T_1]$, $[-T_1,T_1]$, and $[T_1,T]$. In the first segment we turn on the coupling constant $e(t)$ adiabatically. Initially, we have no interactions, so $e(-T) = 0$. At the end of the segment, the coupling constant reaches the desired value $e(-T_1) = e$. As discussed in section \ref{sec:ren}, this value is not the physical value, but is related to it via renormalization. In the weak coupling limit, the relation is given by eq.\ \eqref{eq45}. For $t\in [-T,-T_1]$, we may choose $e^2(t) = \frac{T+t}{T-T_1} e^2$. Similarly, we ought to choose a path for the mass counter term $\delta m$ so that $\delta m(-T) = 0$ and $\delta m (-T_1) = \delta m$, the latter being the desired value. It is determined by renormalization of the mass, and for weak coupling it is given by the perturbative expression \eqref{eqme}, so we may choose $\delta m(t) \approx -1.36 \frac{T+t}{T-T_1} e^2$.

In the second time segment, $t\in [-T_1,T_1]$, the coupling constants are held fixed. Finally, in the third time segment, $t\in [T_1,T]$ the coupling constants are switched off adiabatically following a path which is the reverse path of the one followed during the first time segment.

During time evolution, the unitary operators $e^{i \delta t H_0}$ and $ e^{i \delta t H_{\text{c.t.}}}$ are Gaussian (since the corresponding Hamiltonians, \eqref{eqH0} and \eqref{eqHct} are quadratic in the fields), and can be straightforwardly implemented with a network of optical elements.

The unitary $e^{i\delta t H_I(e)}$ involves interaction terms that are cubic and quartic in the fields (eq.\ \eqref{Hint}). Notice that they only involve the fields and not their conjugate momenta. Since the fields act as coordinate quadratures on a state, they can be implemented by cubic and quartic phase gates, e.g., as detailed in \cite{Marshall2014,Marshall:2015mna,Lau2017}.

After time evolution, at $t=-T$, we \textit{uncompute} the system by applying the Gaussian unitary $\mathcal{U}$ (eq.\ \eqref{eq53}), and measure the photon number of each qumode. These measurements provide us with a distribution of particles (scalar particles and antiparticles, as well as photons) in the final state $|\text{out}\rangle$, on account of \eqref{eq53}. Thus, we obtain the scattering cross section corresponding to the scattering amplitude \eqref{eq56}.

\section{Conclusion}
\label{sec:con}
Quantum computation based on a continuous-variable architecture (qumodes) has received significant attention in recent years due to its experimental advantages over discrete variables (qubits). However, quantum algorithms based on continuous variables are a lot less developed than discrete-variable algorithms. We presented a continuous-variable quantum algorithm for the calculation of scattering amplitudes of massive charged scalars coupled to photons, by extending the quantum algorithm for self-interacting chargeless scalars of ref.~\cite{Marshall:2015mna}. The calculation of scattering amplitudes is one of the most challenging problems in quantum field theory, especially in cases involving a large number of particles which interact strongly. As in the discrete-variable case \cite{Jordan:2011ci}, the quantum algorithm offers an exponential speedup over known classical algorithms based on lattice gauge theory. Thus, such quantum algorithms will allow us to understand particle interactions beyond perturbation theory.

Working with quantum electrodynamics presented a number of complications compared with chargeless scalars emanating from gauge invariance. We showed that gauge invariance could be ensured if one simply imposed it on the non-interacting system and then adiabatically turned on the electric coupling constant. Moreover, our gauge choice led to an interaction Hamiltonian which only depended on the fields and not on their conjugate momenta. Thus, it could be implemented with higher-order (non-Gaussian) phase gates, as in the case of scalar field theory \cite{Marshall:2015mna}. We did not include a scalar self-interaction term, although this can be added straightforwardly. 

It would be interesting to extend our approach to fermionic fields in order to understand the dynamics of electrons, etc. It is also desirable to develop a similar algorithm for non-abelian gauge quantum field theories in order to understand (weak and strong) nuclear forces. Work in this direction is in progress.
\acknowledgments

G.S.\ acknowledges support from the U.S.\ Office of Naval Research under award number N00014-15-1-2646.

%

\begin{appendix}

\end{appendix}


\begin{thebibliography}{00}


\bibitem{Jordan:2011ci} 
  S.~P.~Jordan, K.~S.~M.~Lee and J.~Preskill,
  ``Quantum Computation of Scattering in Scalar Quantum Field Theories,''
  Quantum Information and Computation {\bf 14}, 1014-1080 (2014)
  {\href{https://arxiv.org/abs/1112.4833}{[arXiv:1112.4833 [hep-th]]}}.


%
%
%
%
%
%
%
%
  
  
\bibitem{Lloyd1999} 
  S.~Lloyd,~S.~L.~Braunstein
  ``Quantum Computation over Continuous Variables,''
  {\href{https://doi.org/10.1103/PhysRevLett.82.1784}{Phys. Rev. Lett. {\bf 82}, 1784}} (1999)
  
\bibitem{Pati2000cv} 
  A.~K.~Pati,~S.~L.~Braunstein, and S.~Lloyd,
  ``Quantum Searching with Continuous Variables,''
 {\href{https://arxiv.org/abs/quant-ph/0002082}{[arXiv:quant- ph/0002082]}}(2000).
   
   \bibitem{Lomonaco2003} 
   S. J. Lomonaco Jr. and L. H. Kauffman. 
   ``A Continuous Variable Shor Algorithm," 
  {\href{https://arxiv.org/abs/quant-ph/0210141}{[arXiv:quant-ph/0210141]}} (2002).
  
  \bibitem{Braunstein2003} 
  S.~L.~Braunstein, and A.~K.~Pati,
  ``Quantum Information with Continuous Variables,''
  {\href{http://www.springer.com/us/book/9781402011955}{Kluwer Academic Publisher}} (2003).
  
  \bibitem{Adcock2009}
  M. R. A. Adcock, P. H\o yer, and B. C. Sanders. 
  ``Limitations on Continuous Variable Quantum Algorithms with Fourier Transforms",
   {\href{http://iopscience.iop.org/article/10.1088/1367-2630/11/10/103035/meta}{New J. Phys., {\bf 11}, 103035}} (2009).
  
  \bibitem{Zwierz2010}
 M. Zwierz, C. A. P\'{e}rez-Delgado, and P. Kok, 
 ``Unifying Parameter Estimation and the Deutsch-Jozsa Algorithm for Continuous Variables",
 {\href{https://journals.aps.org/pra/abstract/10.1103/PhysRevA.82.042320}{Phys. Rev. A {\bf 82}, 042320}} (2010)
  
  \bibitem{Braunstein1999} 
  S.~L.~Braunstein, and H.~J. Kimble, 
  ``Dense Coding for Continuous Variables,''
{\href{https://journals.aps.org/pra/abstract/10.1103/PhysRevA.61.042302}{Phys. Rev. A {\bf 61}, 042302}} (2000).
  
  \bibitem{Li2002}
  X. Li, Q. Pan, J. Jing, J. Zhang, C. Xie, and K. Peng,
  ``Quantum Dense Coding Exploiting a Bright Einstein-Podolsky-Rosen Beam",
{\href{https://journals.aps.org/prl/abstract/10.1103/PhysRevLett.88.047904}{Phys. Rev. Lett. {\bf 88}, 047904}} (2002)
  
  \bibitem{Laudenbachs2017}
  F. Laudenbach, C. Pacher, Chi-Hang Fred Fung, et.al.
  ``Continuous-Variable Quantum Key Distribution with Gaussian Modulation - The Theory of Practical Implementations",
  {\href{https://arxiv.org/abs/1703.09278}{[arXiv:1703.09278v2 [quant-ph]] }}(2017)
  
  \bibitem{Grosshans2003}
  F. Grosshans, et al., 
  ``Quantum Key Distribution Using Gaussian-modulated Coherent States,"
   Nature {\bf 421}, 238-41 (2003), {\href{https://arxiv.org/abs/quant-ph/0312016}{[arXiv:quant-ph/0312016]}}.
  
  \bibitem{Aoki2009}
  T.~Aoki,~et.~al.,
  ``Quantum Error Correction Beyond Qubits",
  Nature Physics {\bf 5}, 541 - 546 (2009), {\href{https://arxiv.org/abs/0811.3734}{[arXiv:0811.3734 [quant-ph]]}}. 
  
  \bibitem{Vaidman1994}
  L. Vaidman,
  ``Teleportation of Quantum States",
{\href{https://journals.aps.org/pra/abstract/10.1103/PhysRevA.49.1473}{Phys. Rev. A {\bf 49}, 1473}} (1994)
  
  
    \bibitem{Braunstein1998} 
  S.~L.~Braunstein, and H.~J. Kimble, 
  ``A Posteriori Teleportation,''
{\href{https://www.nature.com/nature/journal/v394/n6696/full/394840a0.html}{Nature 394, 840-841}} (1998).
  
  \bibitem{Andersen2015}
  U. L. Andersen, J. S. Neergaard-Nielsen,	P. v. Loock, and A. Furusawa
  ``Hybrid discrete- and continuous-variable quantum information",
  Nature Physics {\bf 11}, 713-719 (2015), {\href{https://arxiv.org/abs/1409.3719}{arXiv:1409.3719 [quant-ph]}}.
  
  \bibitem{Menucci2006}
 N. C. Menicucci, P. van Loock, M. Gu, C. Weedbrook, T. C. Ralph, and M. A. Nielsen, 
 ``Universal Quantum Computation with Continuous-Variable Cluster States",
 {\href{https://journals.aps.org/prl/abstract/10.1103/PhysRevLett.97.110501}{Phys. Rev. Lett. {\bf 97}, 110501 (2006)}}.
  
  \bibitem{Lau2013}
  H.-K. Lau and C. Weedbrook
  ``Quantum Secret Sharing with Continuous-variable Cluster States",
  {\href{https://journals.aps.org/pra/abstract/10.1103/PhysRevA.88.042313}{Phys. Rev. A {\bf 88}, 042313 (2013)}}.
  
  
  \bibitem{Yokoyama2015}
  S. Yokoyama, R. Ukai, S. C. Armstrong, J.-i. Yoshikawa, P. van Loock, and A. Furusawa, 
  ``Demonstration of a Fully Tunable Entangling Gate for Continuous-variable One-way Quantum Computation",
  {\href{https://journals.aps.org/pra/abstract/10.1103/PhysRevA.92.032304}{Phys. Rev. A {\bf 92}, 032304 (2015)}}.
  
 \bibitem{Miyata2014}
 K. Miyata, H. Ogawa, P. Marek, R. Filip, H. Yonezawa, J.-i. Yoshikawa, and A. Furusawa, 
 ``Experimental Realization of a Dynamic Squeezing Gate",
{\href{https://journals.aps.org/pra/abstract/10.1103/PhysRevA.90.060302}{Phys. Rev. A {\bf 90}, 060302 (2014)}}.


\bibitem{Pysher2011}
M. Pysher, Y. Miwa, R. Shahrokhshahi, R. Bloomer, and O. Pfister, 
``Parallel Generation of Quadripartite Cluster Entanglement in the Optical Frequency Comb",
{\href{https://journals.aps.org/prl/abstract/10.1103/PhysRevLett.107.030505}{Phys. Rev. Lett. {\bf 107}, 030505 (2011)}}.
  

  
  

\bibitem{Marshall:2015mna} 
  K.~Marshall, R.~Pooser, G.~Siopsis and C.~Weedbrook,
  ``Quantum Simulation of Quantum Field Theory Using Continuous Variables,''
  {\href{https://journals.aps.org/pra/abstract/10.1103/PhysRevA.92.063825}{Phys.\ Rev.\ A {\bf 92}, no. 6, 063825}} (2015)
  [arXiv:1503.08121 [quant-ph]].

\bibitem{Martinez2016}
E. A. Martinez, C. A. Muschik, P. Schindler, et.al.,
``Real-time dynamics of lattice gauge theories with a few-qubit quantum computer,"
{\href{http://www.nature.com/nature/journal/v534/n7608/full/nature18318.html?foxtrotcallback=true}{Nature {\bf 534}, 516-519}} (2016).

\bibitem{Yang:2016hjn} 
  D.~Yang, G.~S.~Giri, M.~Johanning, C.~Wunderlich, P.~Zoller and P.~Hauke,
  ``Analog quantum simulation of (1+1)-dimensional lattice QED with trapped ions,''
  {\href{https://journals.aps.org/pra/abstract/10.1103/PhysRevA.94.052321}{Phys.\ Rev.\ A {\bf 94}, no. 5, 052321}} (2016)
  [arXiv:1604.03124 [quant-ph]].

\bibitem{Notarnicola:2015sia} 
  S.~Notarnicola, E.~Ercolessi, P.~Facchi, G.~Marmo, S.~Pascazio and F.~V.~Pepe,
  ``Discrete Abelian Gauge Theories for Quantum Simulations of QED,''
 {\href{http://iopscience.iop.org/article/10.1088/1751-8113/48/30/30FT01/meta;jsessionid=2E00D9F21F48800F7F5FF55D79154724.ip-10-40-1-105}{ J.\ Phys.\ A {\bf 48}, no. 30, 30FT01}} (2015)
  [arXiv:1503.04340 [quant-ph]].

\bibitem{Ercolessi:2017jbi} 
  E.~Ercolessi, P.~Facchi, G.~Magnifico, S.~Pascazio and F.~V.~Pepe,
  ``Quantum Simulation of QED in 1D: Evidence of a Phase Transition,''
{\href{https://arxiv.org/abs/1705.11047}{arXiv:1705.11047 [quant-ph]}} (2017).
  
\bibitem{MarkSre}
  M.~Srednicki,
``Quantum Field Theory'',
{\href{http://www.physics.ucsb.edu/~mark/qft.html}{Cambridge University Press }}(2007).
  
\bibitem{Weedbrook2012}
C.~Weedbrook, S.~Pirandola, R.~Garc\'{\i}a-Patr\'on, N.~J.~Cerf, T.~C.~Ralph, J.~H.~ Shapiro, and S.~Lloyd, ``Gaussian Quantum Information", 
{\href{https://journals.aps.org/rmp/abstract/10.1103/RevModPhys.84.621}{Rev. Mod. Phys. {\bf{84}}, 2, 621-669}} (2012).


\bibitem{Kitaev2002}
A. Yu Kitaev, A.H. Shen, and M.N. Vyalyi, 
``Classical and Quantum Computation," 
{\href{http://www.ams.org/books/gsm/047/}{American Mathematical Society}} (2002)


\bibitem{Reck1994}
M.~Reck, A.~Zeilinger, H.~J.~Bernstein, P.~Bertani,
``Experimental Realization of any Discrete Unitary Operator'', 
{\href{https://journals.aps.org/prl/abstract/10.1103/PhysRevLett.73.58}{Phys. Rev. Lett. {\bf{73}}, 58}} (1994).

\bibitem{Braunstein2005}
S.~L.~Braunstein,
``Squeezing as an Irreducible Resource'', 
{\href{https://journals.aps.org/pra/abstract/10.1103/PhysRevA.71.055801}{Phys.\ Rev.\ A {\bf{71}},~055801}} (2005).

\bibitem{bibadia}
J. E. Avron and A. Elgart, ``Adiabatic theorem without a gap condition: Two-level system coupled to quantized radiation field," {\href{https://journals.aps.org/pra/abstract/10.1103/PhysRevA.58.4300}{Phys.\ Rev.\ A \textbf{58}, 4300}} (1998).

\bibitem{Hong1986}
C. K. Hong and L. Mandel,
``Experimental Realization of a Localized One-Photon State",
{\href{https://journals.aps.org/prl/abstract/10.1103/PhysRevLett.56.58}{Phys.\ Rev.\ Lett.\ {\bf 56}, 58-60}} (1986)

\bibitem{Migdall2013}
A. Migdall, S. V. Polyakov, J. Fan and J. C. Bienfang,
``Single-Photon Generation and Detection Physics and Applications",
{\href{http://www.sciencedirect.com/science/bookseries/10794042/45}{Elsevier Science, Experimental Methods in the Physical Sciences,  {\bf 45}, 1-562}} (2013)

%
%
%


\bibitem{Braunstein2004}
S.\ L.\ Braunstein and P.\ van Loock, 
``Quantum Information with Continuous Variables",
{\href{https://journals.aps.org/rmp/abstract/10.1103/RevModPhys.77.513}{Rev.\ Mod.\ Phys.\ {\bf{77}}, 513}} (2005).

\bibitem{Marshall2014}
K.\ Marshall, R.\ Pooser, G.\ Siopsis, and C.\ Weedbrook, 
``Repeat-until-success cubic phase gate for universal continuous-variable quantum computation,"
{\href{https://journals.aps.org/pra/abstract/10.1103/PhysRevA.92.063825}{Phys.\ Rev.\ A {\bf 91}, 032321}} (2015).

\bibitem{Lau2017}
H.-K.\ Lau, R.\ Pooser, G.\ Siopsis, and C.\ Weedbrook,
``Quantum Machine Learning over Infinite Dimensions," 
{\href{https://journals.aps.org/prl/abstract/10.1103/PhysRevLett.118.080501}{Phys.\ Rev.\ Lett.\ {\bf{118}}, 080501}} (2017).

%



\end{thebibliography}
\end{document}